\documentclass[12pt]{article}
\usepackage{graphicx,epsfig}
\usepackage{color}
\usepackage{cite}
\title{Combined effects of local and nonlocal hybridization
on formation and condensation of excitons in the extended
Falicov-Kimball model}
\author{Pavol Farka\v sovsk\'y\\
Institute  of  Experimental  Physics,  Slovak   Academy   of
Sciences\\
Watsonova 47, 043 53 Ko\v {s}ice, Slovakia}
\date{}
\begin{document}
\baselineskip=24pt
\maketitle

\begin{abstract}
We study the combined effects of local and nonlocal hybridization on the 
formation and condensation of the excitonic bound states in the extended 
Falicov-Kimball model by the density-matrix-renormalization-group 
(DMRG) method. Analysing the resultant behaviours of the  excitonic 
momentum distribution $N(q)$ we found, that unlike the local
hybridization $V$, which supports the formation of the $q=0$ momentum
condensate, the nonlocal hybridization $V_n$ supports the formation of
the $q=\pi$ momentum condensate. The combined effect 
of local and nonlocal hybridization further enhances the excitonic 
correlations in $q=0$ as well as $q=\pi$ state, especially for $V$ 
and $V_n$ values from the charge-density-wave (CDW) region. Strong 
effects of local and nonlocal hybridization are observed also 
for other ground-state quantities of the model such as the  
$f$-electron density, or the density of unbound $d$-electrons,
which are generally enhanced with increasing $V$ and $V_n$.
The same calculations performed for nonzero values of $f$-level
energy $E_f$ revealed that this model can yield a reasonable explanation 
for the pressure-induced resistivity anomaly observed experimentally 
in $TmSe_{0.45}Te_{0.55}$ compound. 

\end{abstract}
%\thanks{PACS nrs.: 71.27.+a, 71.28.+d, 71.30.+h}

\newpage
\section{Introduction}
The formation of excitonic quantum condensates is an intensively
studied continuous problem in condensed matter physics~\cite{Blatt,Keldysh,
Mos,Litt}.
Whilst theoretically predicted a long time ago~\cite{theo}, 
no conclusive experimental proof of the existence of the excitonic 
condensation has been achieved yet. However, the latest 
experimental studies of materials with strong electronic correlations 
showed that promising candidates for the experimental 
verification of the excitonic condensation could be 
$TmSe_{0.45}Te_{0.55}$~\cite{Wachter},$1T-TiSe_2$~\cite{Monney},
$Ta_2NiSe_5$~\cite{Wakisaka}, or a double bilayer graphene system
~\cite{Perali}. In this regard, the mixed valence compound
$TmSe_{0.45}Te_{0.55}$ was argued to exhibit a pressure-induced 
excitonic instability, related to an anomalous increase in the 
electrical resistivity~\cite{Wachter}. In particular, the detailed studies 
of the pressure-induced semiconductor-semimetal transition in this 
material (based on the Hall effect, electrical and thermal (transport)
measurements) showed that excitons are created in a large number and condense 
below 20~K. On the other hand, in the layered transition-metal dichalcogenide 
$1T-TiSe_2$, a BCS-like electron-hole pairing was considered as the driving 
force for the periodic lattice distorsion~\cite{Monney}. 
Moreover, quite recently, the excitonic-insulator state was probed 
by angle-resolved photoelectron spectroscopy in the semiconducting 
$Ta_2NiSe_5$ compound~\cite{Wakisaka}. At present, it is generally accepted 
that the minimal theoretical model for a description of excitonic 
correlations in these materials could be the Falicov-Kimball
model~\cite{Falicov} and its extensions~\cite{B1,B2,Z1,Phan,
Seki,Z2,Kaneko1,Kaneko2,Ejima,Fark1}. The  original Falicov-Kimball model 
describes a tight-binding system of itinerant $d$ electrons 
interacting via the on-site Coulomb repulsion with localized $f$
electrons (the spin degrees of freedom of the
$d$ and $f$ electrons are not included):
\begin{equation}
H_0=-t_d\sum_{\langle i,j \rangle}d^+_id_j+U\sum_if^+_if_id^+_id_i+E_f\sum_if^+_if_i,
\end{equation}
where $f^+_i$, $f_i$ are the creation and annihilation
operators  for an electron in  the localized state at 
lattice site $i$ with binding energy $E_f$ and $d^+_i$,
$d_i$ are the creation and annihilation operators of the itinerant 
spinless electrons (with the nearest-neighbor $d$-electron
hopping constant $t_d$) in the $d$-band Wannier state at site $i$.
In what follows we consider $t_d=1$ and all energies
are measured in units of $t_d$.

Since the local $f$-electron  number $f^+_if_i$ is strictly
conserved quantity, the $d$-$f$ electron coherence cannot be
established in this model. One way to overcome this shortcoming  
is to include an explicit local hybridization 
$H_V=V\sum_id^+_if_i+f^+_id_i$ between the $d$
and $f$ orbitals. This model has been extensively
studied in our previous work~\cite{Fark1}.
The numerical analysis of the excitonic momentum distribution 
$N(q)=\langle b^+_qb_q\rangle$ (with $b^+_q=(1/\sqrt{L})\sum_k
d^+_{k+q}f_k$, where $L$ denotes the number of lattice sites)
showed that this quantity diverges for $q=0$, signalizing a
massive condensation of preformed excitons at this momentum.
The stability of the zero-momentum ($q=0$) condensate
against the $f$-electron hopping has been studied in our very recent 
paper~\cite{Fark_prb}. It was found that the negative values of the $f$-electron 
hopping integrals $t_f$ support the formation of zero-momentum condensate, 
while the positive values of $t_f$ have the fully opposite effect. Moreover,
it was shown that the fully opposite effects on the formation of condensate 
exhibit also the local and nonlocal hybridization with an inversion symmetry. 
The first one strongly supports  the formation of condensate, while the second one 
destroys it completely. However, in the real $d$-$f$ systems, the on-site
hybridization $V$ is usually forbidden for parity reasons~\cite{Czycholl}, 
and therefore the fact
that the nonlocal hybridization with an inversion symmetry does not support 
the formation of excitonic condensate strongly limits the class of materials, 
where this phenomenon can be observed. In this situation, the most
promising candidates for studying this phenomenon seem to be the systems 
with equal parity orbitals, where the nonlocal hybridization $H_n$ can 
be written as: 
\begin{equation}
H_n=V_n\sum_{\langle i,j \rangle}(d^+_if_j+H.c.).
\end{equation}
In such systems, the local hybridization $V$ is allowed,
and thus one can examine the combined effects of the local and nonlocal 
hybridization within the unified picture. 
The weak ($U<1$) and strong ($V \ll U$ and $V_n \ll U$) coupling 
limits of the model Hamiltonian $H_0+H_V+H_n$  have been analyzed 
recently by Zenker et al. in~\cite{Vn}, and the corresponding mean-field 
quantum phase diagrams were presented as functions of the model 
parameters $U, V, V_n$ and $E_f$ for the half-filed band case 
$n_f+n_d=1$ and $D=2$. Moreover, examining effects of the local 
$V$ and nonlocal $V_n$ hybridization they found that in the 
pseudospin space ($c^{+}_{i\uparrow}=d^+_i$,$c^{+}_{i\downarrow}=f^+_i$) 
the nonlocal hybridization $V_n$ favors the staggered Ising-type ordering 
allong the $x$ direction, while $V$ favors a uniform polarization 
along the $x$ direction and the staggered Ising-type ordering  
along the $y$ direction.

In the current paper we examine model for arbitrary $V$ and 
$V_n$ and unlike the paper of Zenker et al.~\cite{Vn}  we focus 
our attention primarily on a description of process of formation 
and condensation of exitonic bound states. For this reason
we calculate (by the DMRG method) various ground state characteristics
of the model such as the excitonic momentum distribution 
$N(q)$, the density of zero momentum excitons $n_0=\frac{1}{L}N(q=0)$,
the total exciton density $n_T=\frac{1}{L}\sum_q N(q)$, the total 
$d$-electron density $n_d$ and the total density of unbound $d$ electrons 
$n^{un}_d=n_d-n_T$,  and analyze their bahaviours as functions 
of the local/nolocal hybridization and the $f$-level position $E_f$.     
It should be noted that such a study could be very valuable from 
the point of view of real materials, since taking into account the parametrization 
between the external pressure and the position of the $f$ level ($E_f \sim p $),
one could deduce from the $E_f$ dependencies of the ground state 
characteristics also their $p$ dependencies, at least qualitatively~\cite{Gon}.
As shown at the end of this paper a simple model based on the above
mentioned parametrization can yield a reasonable explanation 
for the pressure-induced resistivity anomaly observed 
experimentally in $TmSe_{0.45}Te_{0.55}$ compound~\cite{Wachter}.

\section{Results and discussion}
\subsection{DMRG results}

To examine combined effects of the local $V$ and nonlocal $V_n $ hybridization
on the formation and condensation of excitonic bound states in the extended 
Falicov-Kimball model we have performed exhaustive DMRG studies 
of the model Hamiltonian $H=H_0+H_V+H_{n}$ for a wide range 
of the model parameters $V,V_n$ and $E_f$ at the total electron density
$n=n_d+n_f=1$ (the half-filled band case). In all examined cases  
we typically keep up to 500 states per block, although in the numerically 
more difficult cases, where the DMRG results converge slower, 
we keep up to 1000 states. Truncation errors~\cite{White}, given by 
the sum of the density matrix eigenvalues of the discarded states, 
vary from $10^{-6}$ in the worse cases to zero in the best cases.

Let us start a discussion of our numerical results for the case 
$E_f=0$. The typical examples of the excitonic momentum distribution $N(q)$
calculated for the represenative values of $V,V_n$ and $U$
are summarized in Fig.~1, where different panels correspond to: 
(a) $V=0$ and  $V_n \ge 0$, (b) $V_n=0$ and $V \ge 0$, 
(c) $V > 0$ and $V_n \ge 0$.
\begin{figure}[h!]
\begin{center}
\includegraphics[width=7.5cm]{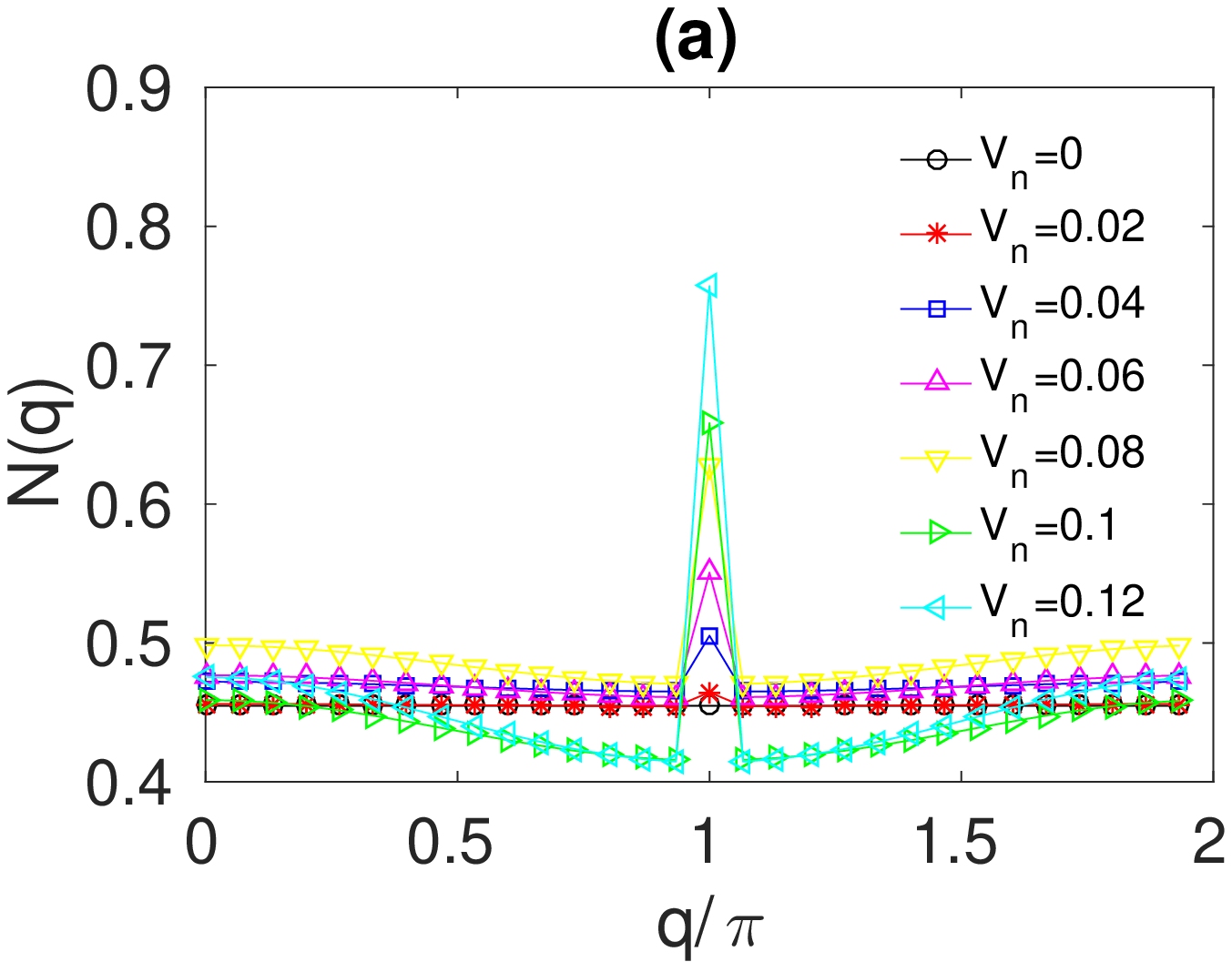}
\includegraphics[width=7.5cm]{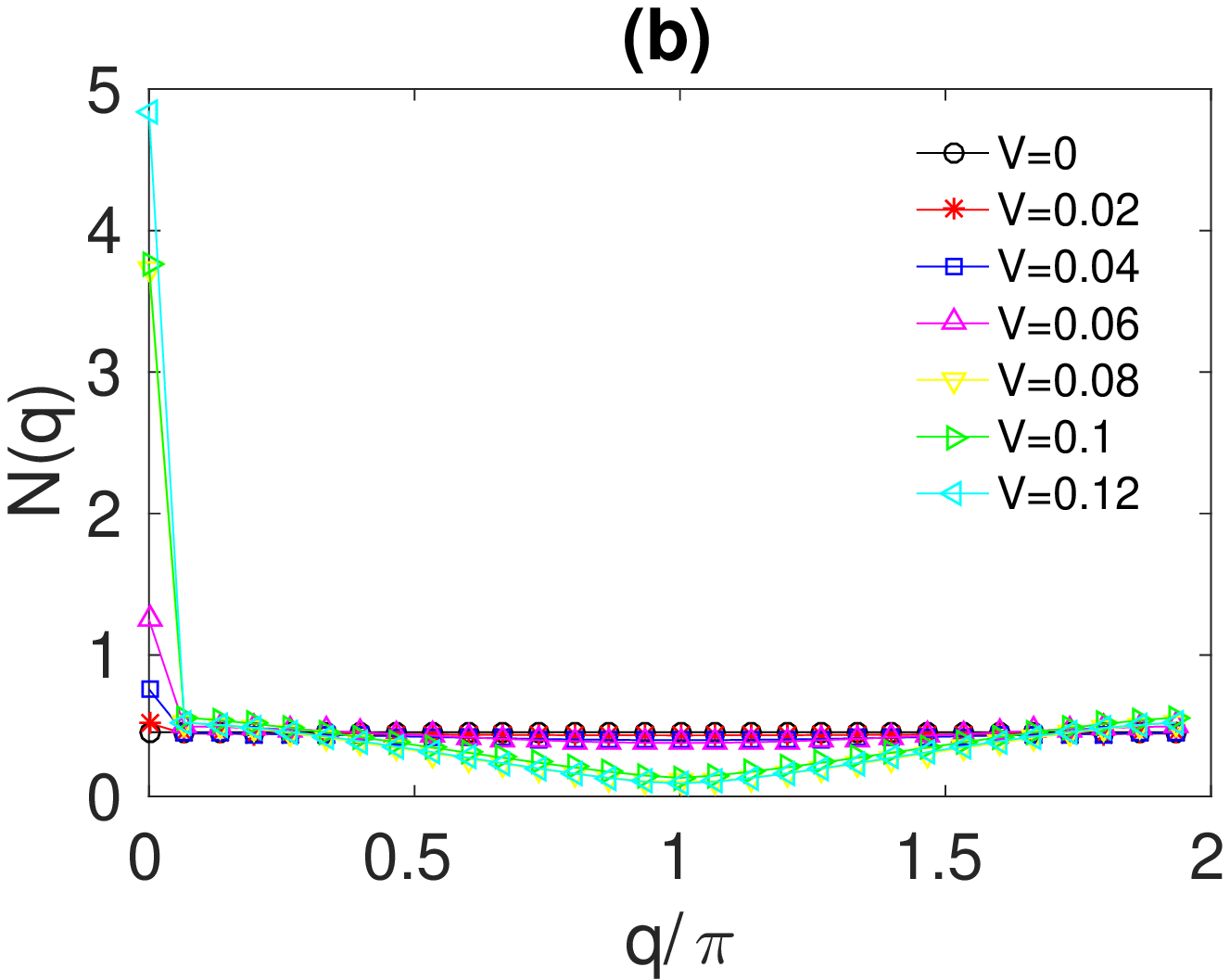}
\includegraphics[width=7.5cm]{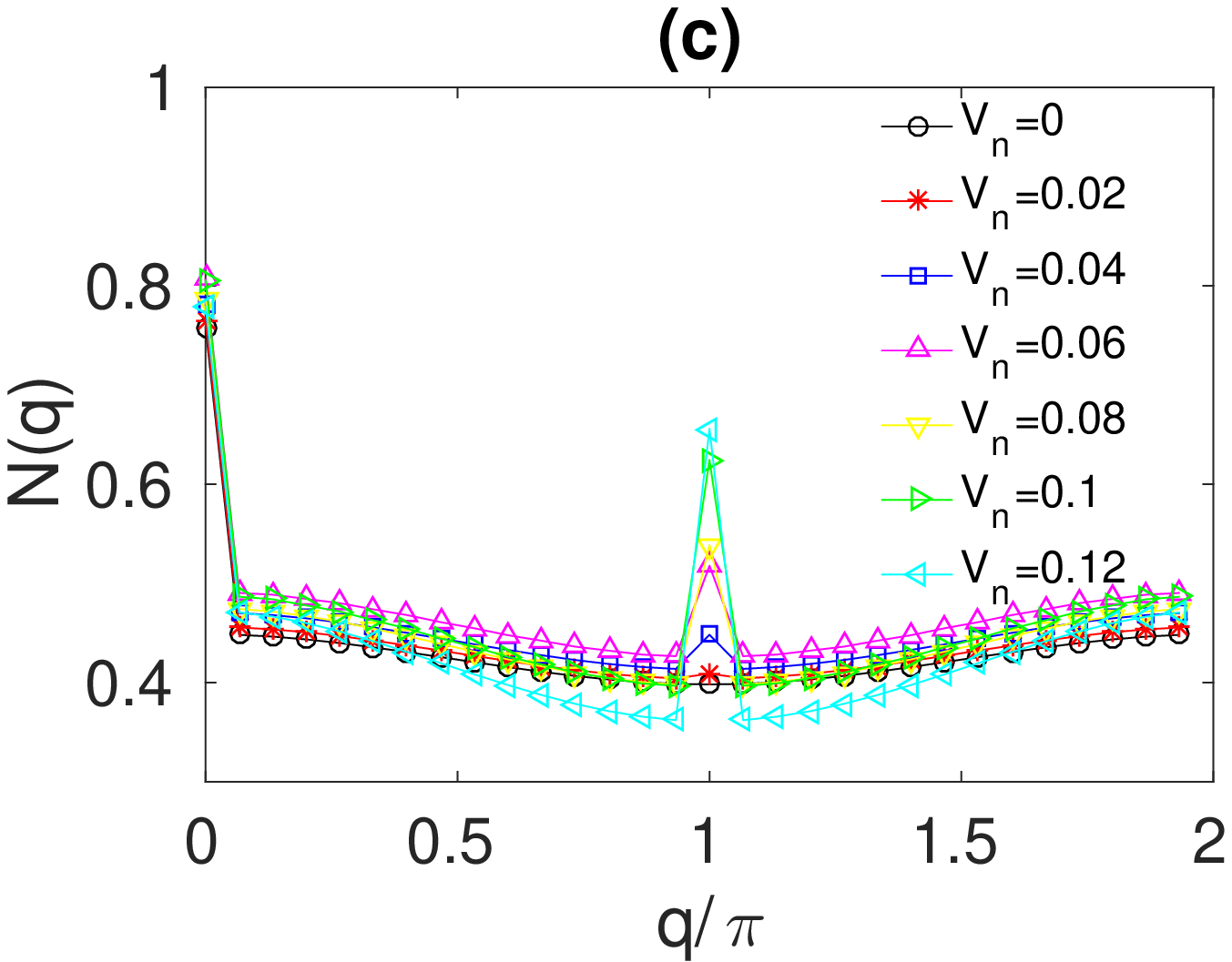}
\includegraphics[width=7.5cm]{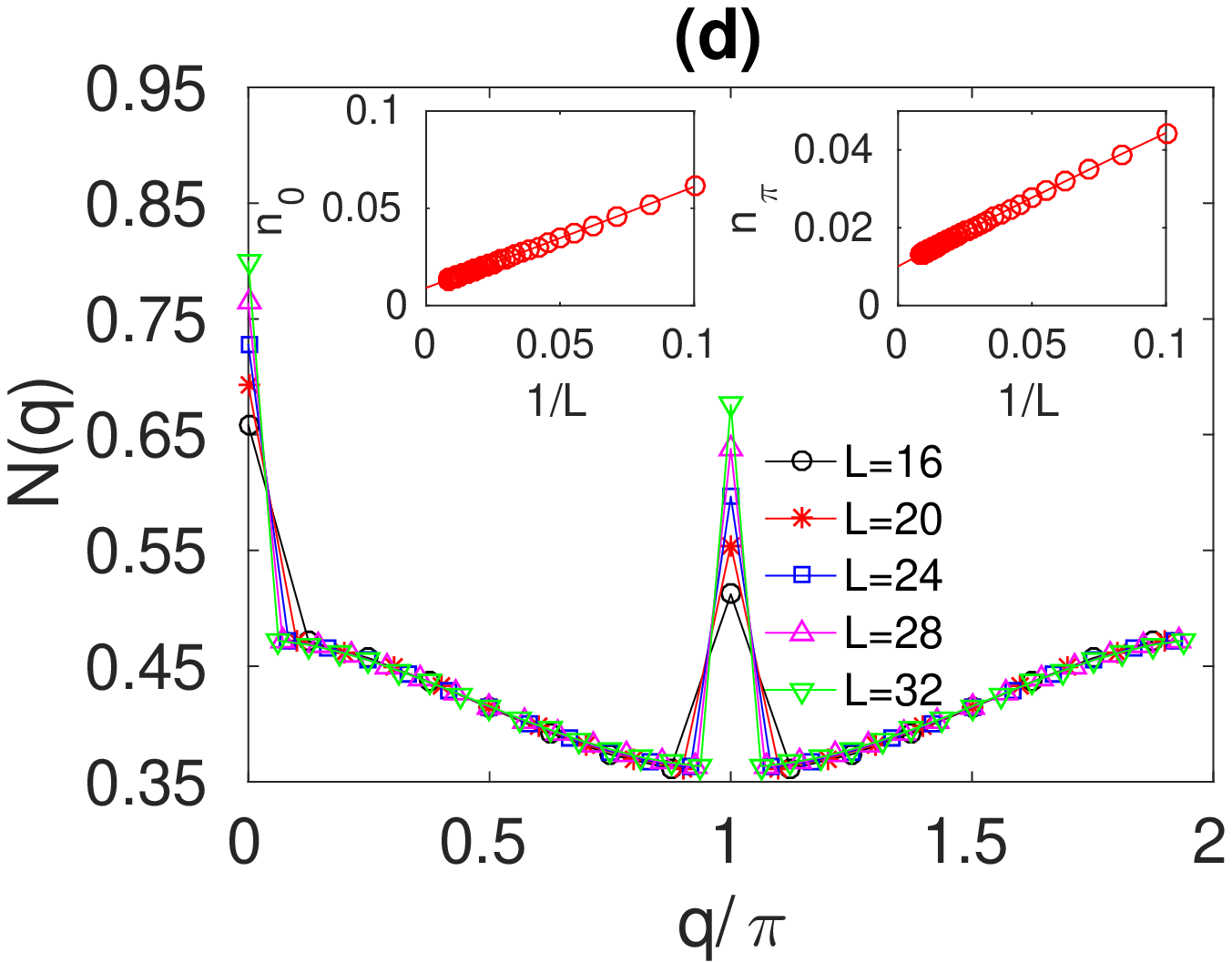}
\end{center}
\vspace*{-0.8cm}
\caption{\small The DMRG results for the excitonic momentum distribution 
$N(q)$ calculated for different model parameters. 
(a) $N(q)$ calculated for $V=0, U=4, E_f=0, L=30$ and different values of $V_n$.
(b) $N(q)$ calculated for $V_n=0, U=4, E_f=0, L=30$ and different values of $V$.
(c) $N(q)$ calculated for $V=0.04, U=4, E_f=0, L=30$ and different values of
$V_n$.
(d) $N(q)$ calculated for $V=0.04, V_n=0.12, U=4, E_f=0$ and different values of
$L$. The insets show the densities of $q=0$ and $q=\pi$ momentum excitons
as functions od $1/L$ calculated for the same values of model parameters.}
\label{figure1}
\end{figure}
Comparing these behaviours one can see that the local and nonlocal hybridization 
exhibit fundamentally different effects on the formation and condensation of excitonic 
bound states. While the local hybridization supports the formation of the 
zero-momentum (ferroelectric) condensate, the nonlocal hybridization supports 
the formation of the $\pi$-momentum (antiferroelectric) condensate.
This is supported by the finite size scaling analysis of the excitonic
momentum distribution $N(q)$ and the densities of $q=0$ and $q=\pi$ momentum
excitons in Fig.~1d. With the increasing cluster size the excitonic momentum 
distribution $N(q)$ diverges at $q=0$ as well as $q=\pi$ 
and both, the density of the zero-momentum condensate ($n_0=N(0)/L$)
as well as  the density of the $\pi$-momentum condensate ($n_{\pi}=N(\pi)/L$)
have the finite values at singular points of $N(q)$ (see the insets in
Fig.~1d). 

Let us now discuss the combined effect of local and nonlocal 
hybridizations on the other ground state quantities of the model.
To describe, in more detail, the process of formation of excitonic 
bound states for nonzero $V$ and $V_n$, we have plotted in Fig.~2, 
the density of zero momentum excitons $n_0$, the density of $q=\pi$ momentum 
excitons $n_\pi$, the total $d$-electron density $n_d$ and the total 
density of unbound $d$ electrons $n^{un}_d=n_d-n_T$ as functions of nolocal 
hybridization $V_n$ for $U=4, E_f=0$ and several different values of $V$. 
\begin{figure}[h!]
\begin{center}
\includegraphics[width=7.5cm]{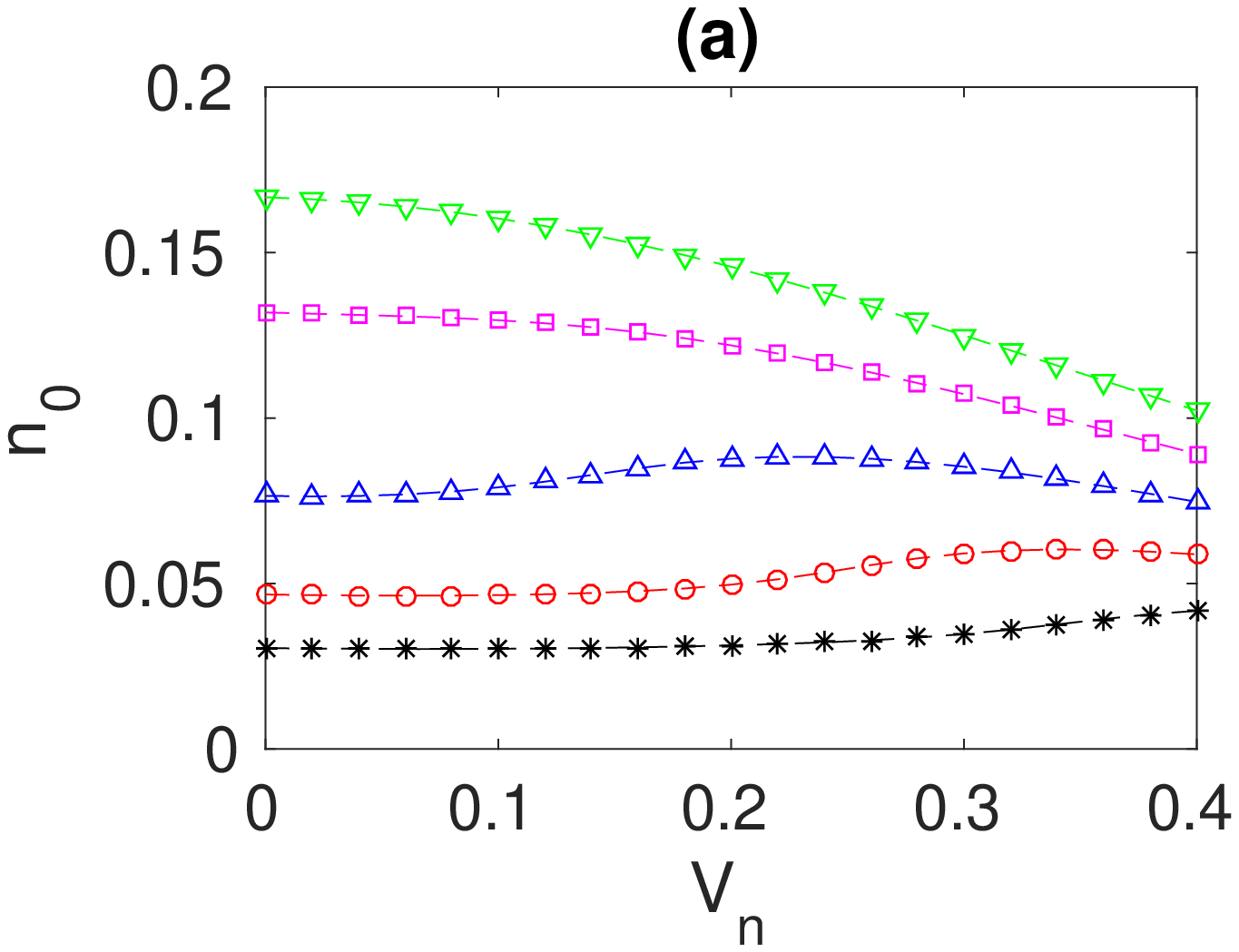}
\includegraphics[width=7.5cm]{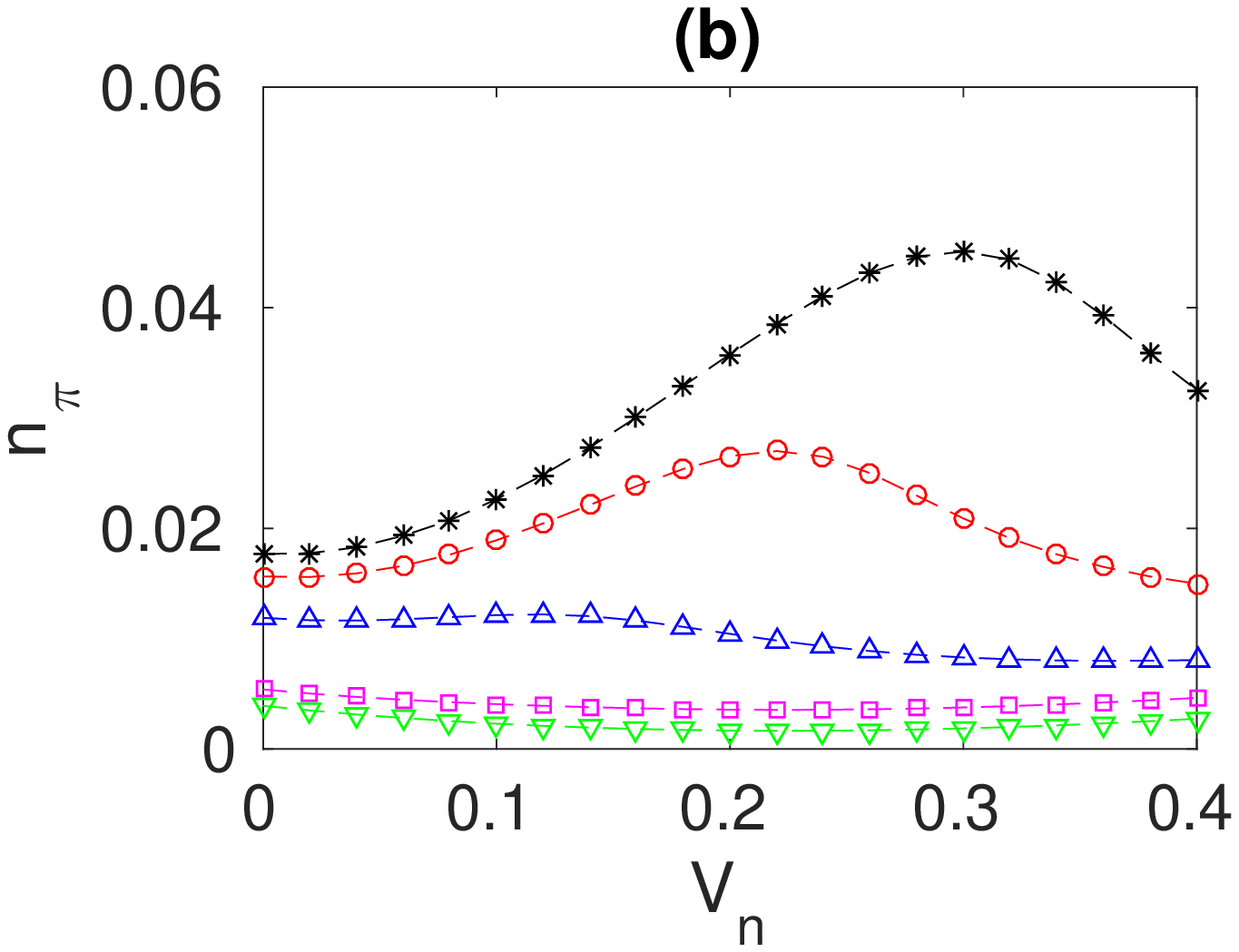}
\includegraphics[width=7.5cm]{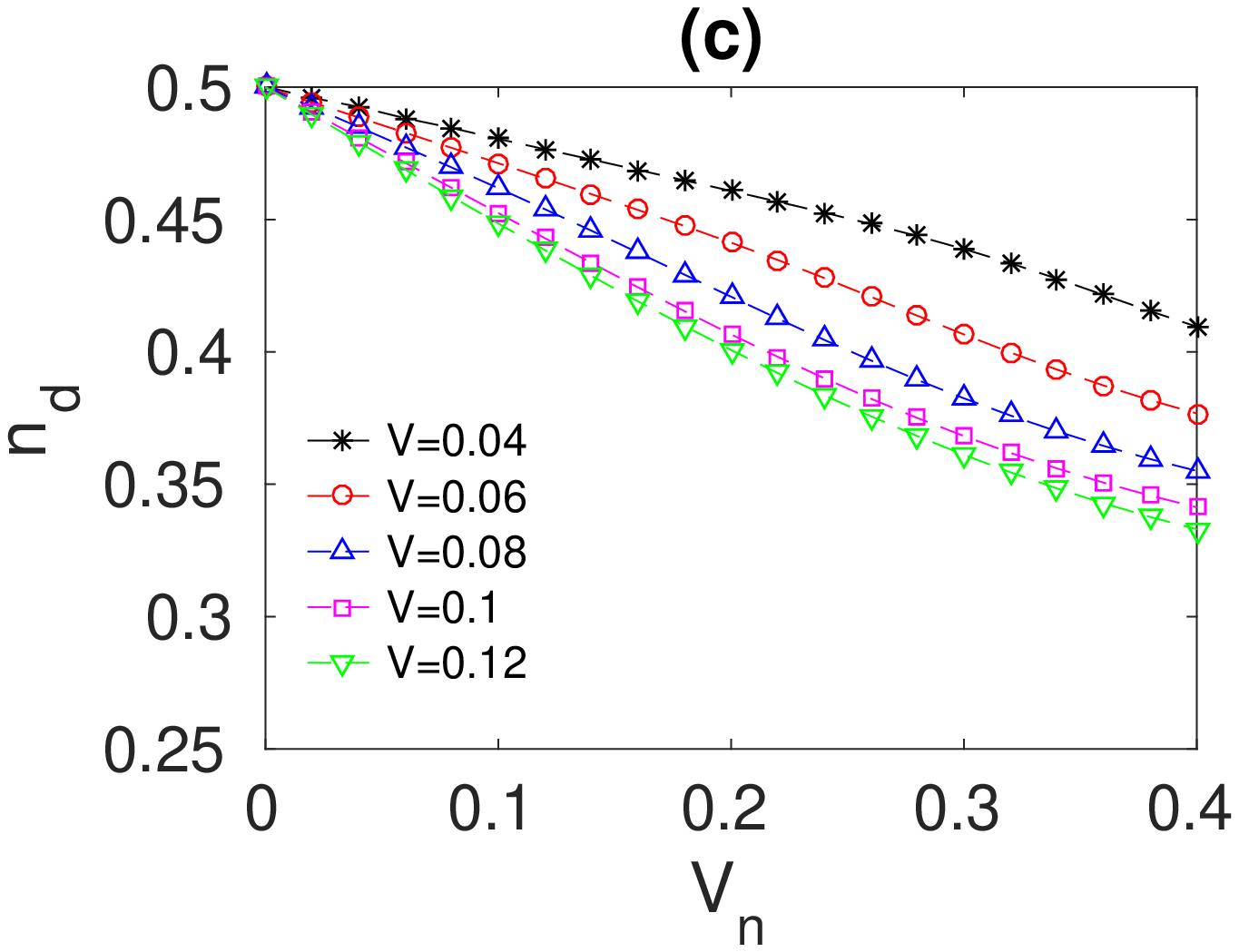}
\includegraphics[width=7.5cm]{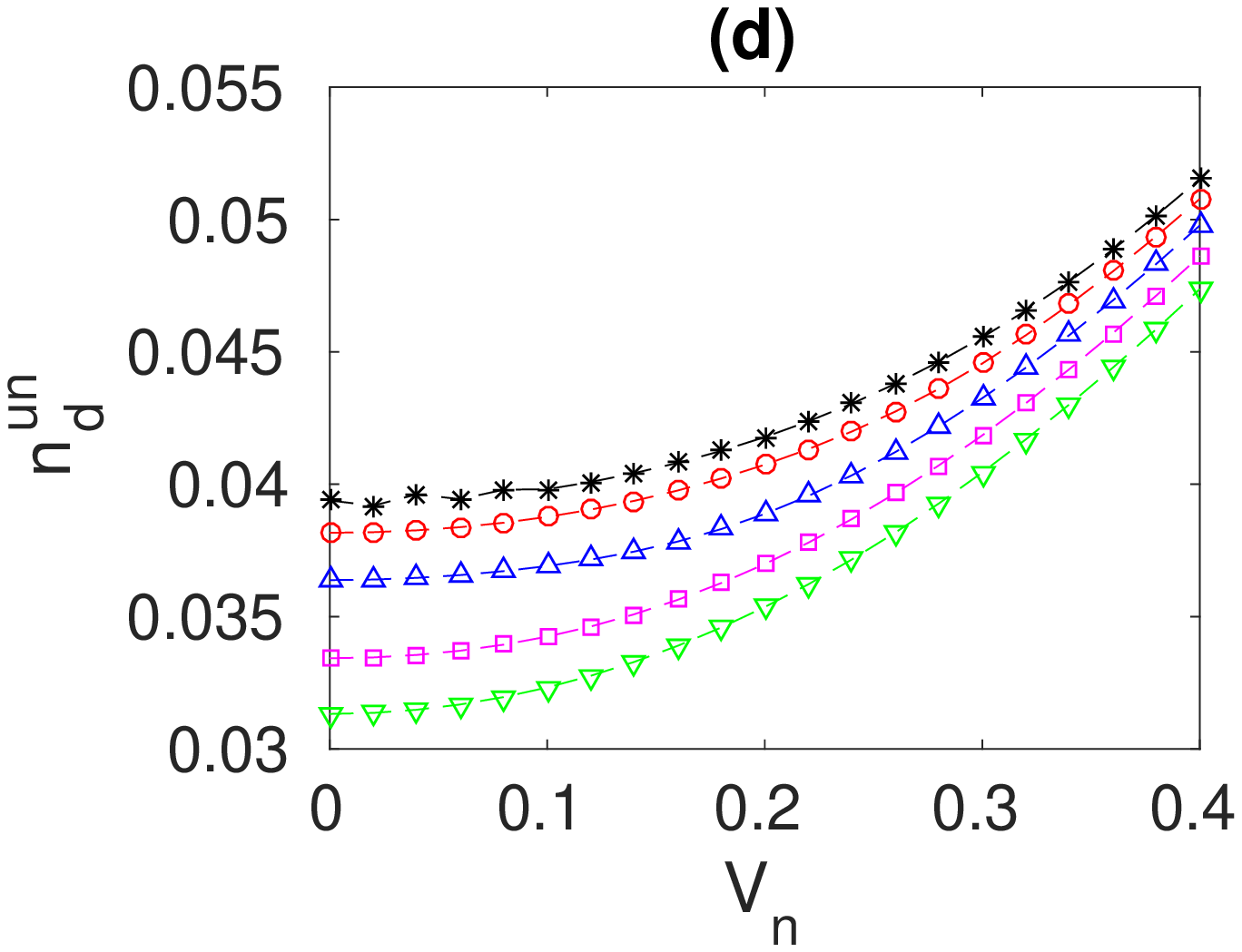}
\end{center}
\vspace*{-0.8cm}
\caption{\small The density of zero-momentum excitons $n_0$ (a), 
the density of $\pi$-momentum excitons $n_{\pi}$ (b),
the total $d$-electron density $n_d$ (c), and the total density of unbound 
$d$ electrons $n^{un}_d=n_d-n_T$ (d) as functions of $V_n$ 
calculated for $U=4, E_f=0, L=60$ and five different values of $V$.}
\label{figure2}
\end{figure}
Fig.~2a shows  the dependence of the density of zero-momentum excitons 
$n_0$ on $V_n$. One can see that the density of the zero-momentum excitons 
$n_0$ as a function of nonlocal hybridizations $V_n$ exhibits two different 
types of behaviours. While for sufficiently large values of $V$ ($V>0.1$), 
the number of zero-momentum excitons is suppressed with the increasing
values of nonlocal hybridization $V_n$, in the opposite limit $V<0.1$,
the number of zero-momentum excitons increases for nonlocal hybridizations
smaller than some critical nonlocal hybridization $V^{c}_n$ and decreases
above this value. To reveal the nature of different behaviour of $n_0(V_n)$ 
below and above $V$ we have calculated numerically the Fourier transform 
of the f-electron density-density correlation function S(q) defined by
\begin{equation}
S(q)=\frac{1}{L}\sum_{jl}e^{iql}(\langle n^f_jn^f_{j+l} \rangle-
\langle n^f_j \rangle \langle n^f_{j+l} \rangle).
\end{equation}
In the regime of long-range order $S(q=\pi) \equiv S(\pi)$ scales linearly
with $L$ and above this regime $S(\pi)$ undergoes a rapid change (see Fig.3a),   
that indicates the transition from the ordered to homogeneous phase.
\begin{figure}[h!]
\begin{center}
\includegraphics[width=7.5cm]{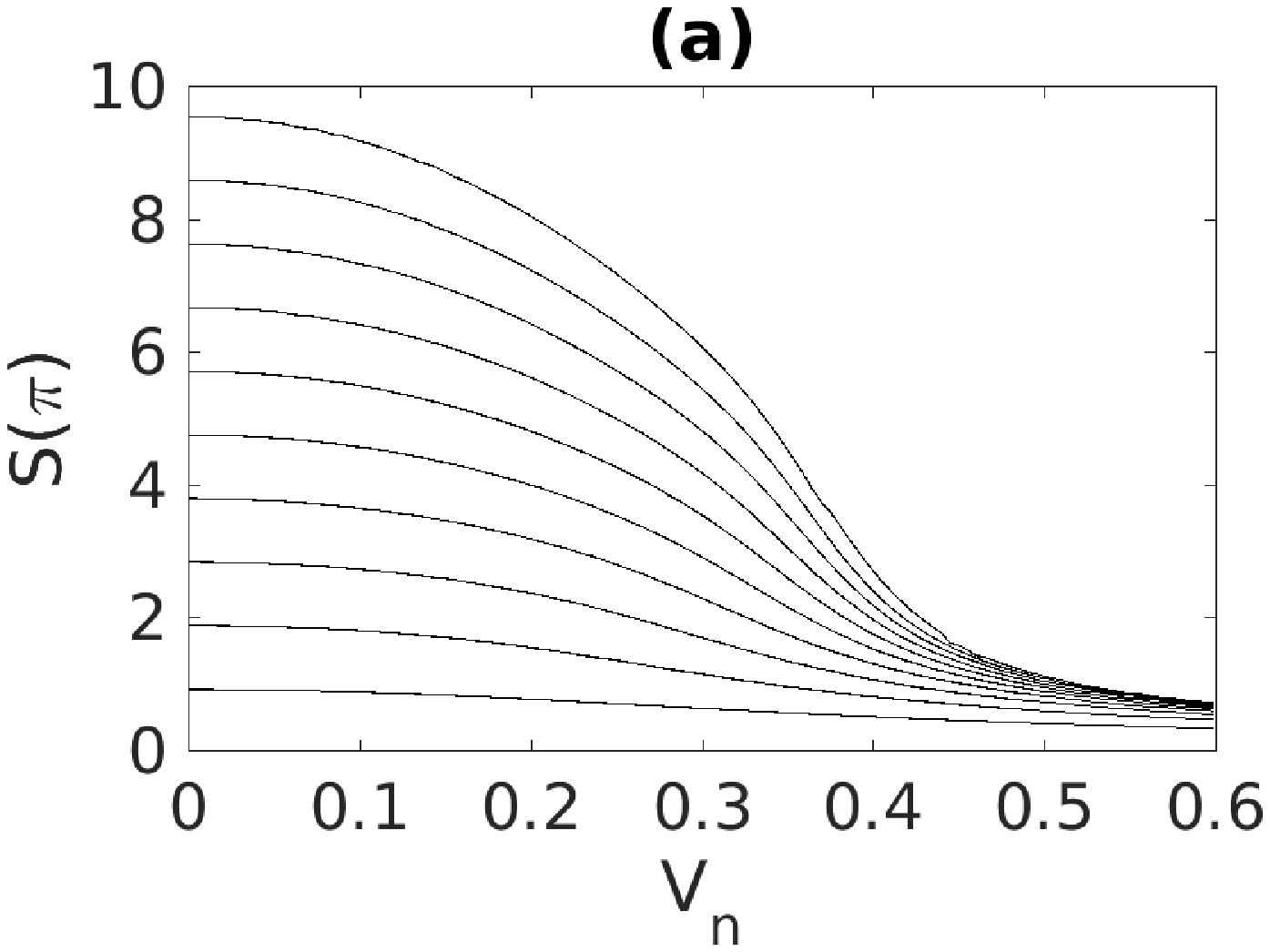}
\includegraphics[width=7.5cm]{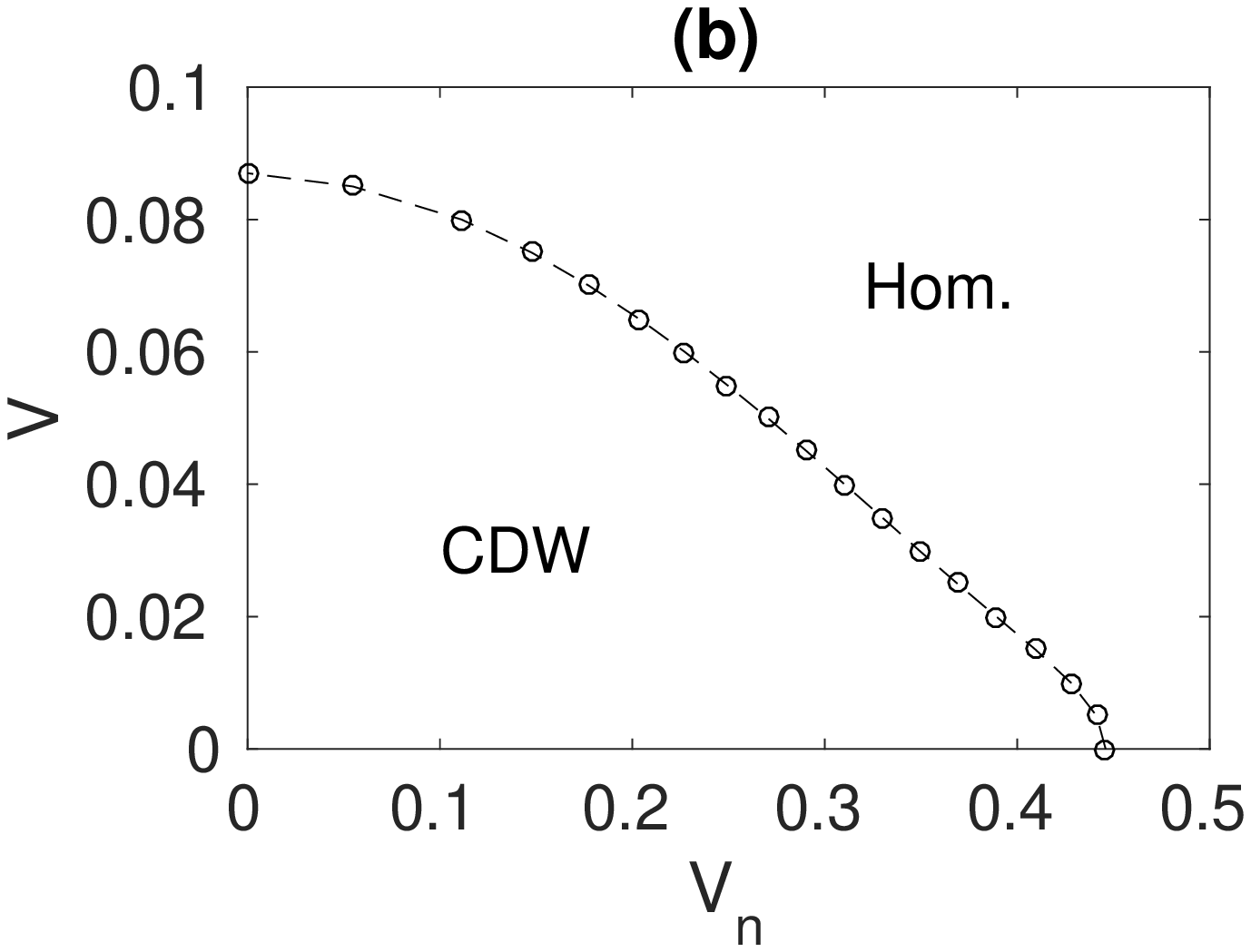}
\end{center}
\vspace*{-0.8cm}
\caption{\small a) The Fourier transform of the $f$-electron density-density 
correlation function $S(\pi)$ as a function of nonlocal hybridization $V_n$, 
calculated for $U=4, E_f=0, V=0.04$ and several different values 
of $L$ (from bottom to top L=4,8,12 \dots 40). 
b) The ground-state phase diagram of the model
in the $V_v$-$V$ plane calculated for $U=4, E_f=0, L=\infty$.
Two different regions correspond to the charge density wave (CDW)
and homogeneous (Hom.) phase.}
\label{figure2}
\end{figure}
Following the procedure described in detail in our previous work~\cite{Fark1},
we have used $S(\pi)$ to identify the phase boundary between the 
long-range order (CDW) state and the homogeneous phase in the $V_n$-$V$ plane.
The complete phase diagram of the model obtained for $U=4$
is displayed in Fig.~3b and it clearly demonstrates that different
behaviour of of $n_0(V_n)$  below and above $V$ is obviously caused
by the CDW order, which supports formation of the zero-momentum condensate.
The CDW order is apparently responsible also for an anomalous increase 
in the density of $q=\pi$ momentum excitons observed for $V$ and $V_n$ 
small (see Fig.~2b) 

The strong combined effects of the local and nonlocal hybridization 
are observed also for the $d$-electron density as a function 
of $V_n$ (see Fig.~2c). While for $V=0$ the $d$-electron density does not
depend on the nonlocal hybridization $V_n$ ($n_d=0.5$ for all $V_n$),
for finite values of local hybridization the $d$-electron density
is considerably reduced by the increasing nonlocal hybridization.    
Since the $f$-electron density $n_f$ is simply given by $n_f=1-n_d$
one obtains the nonlocal hybridization induced valence transitions
even for the case $E_f=0$, that in the absence of nonlocal hybridization
corresponds to the half-filed band case $n_f=n_d=1/2$. 
A very important quantity that characterizes the ground state
of the model is the density of unbound electrons $n^{un}_d=n_d-n_T$.
Its behaviour is shown in Fig.~2d and one can see that this quantity
is considerably enhanced by the nonlocal hybridization. 
 
However, from the physical point of view, it is the most interesting 
to examine changes of $n_0, n_{\pi}, n_d$ and $n^{un}_d$ as functions 
of the $f$-level position $E_f$, since taking into account the parametrization 
between the external pressure $p$ and the $f$-level energy $E_f$~\cite{Gon},
such a study can give us the answer to the very important question,
and namely, how these quantities change with applied pressure $p$. 
In Fig.~4 we present the resultant behaviours of $n_0, n_{\pi}, n_d, n^{un}_d$ 
as functions of the $f$-level position $E_f$ obtained by the DMRG method for 
$V=0.2$ and several different values of $V_n$. 
\begin{figure}[h!]
\begin{center}
\includegraphics[width=7.5cm]{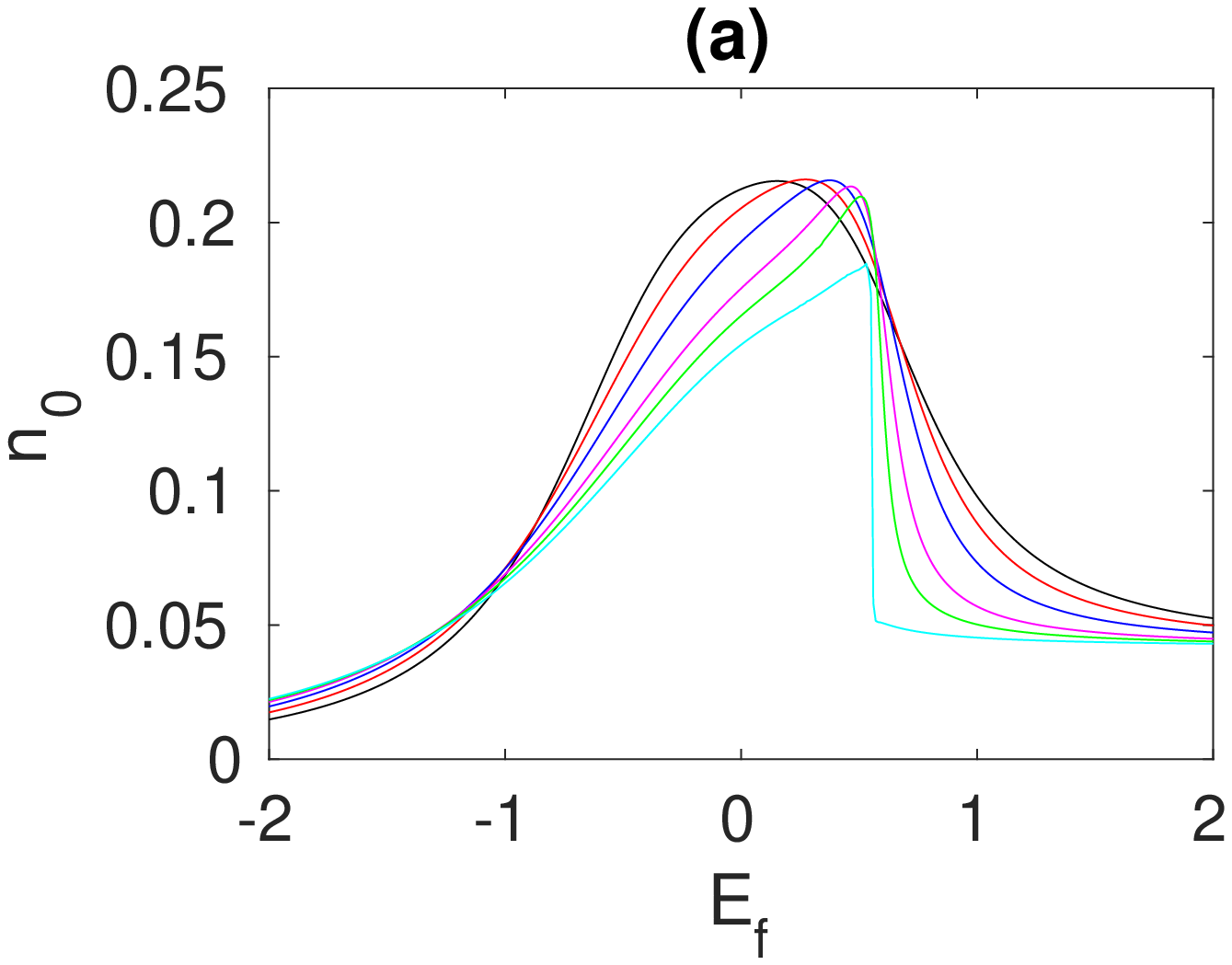}
\includegraphics[width=7.5cm]{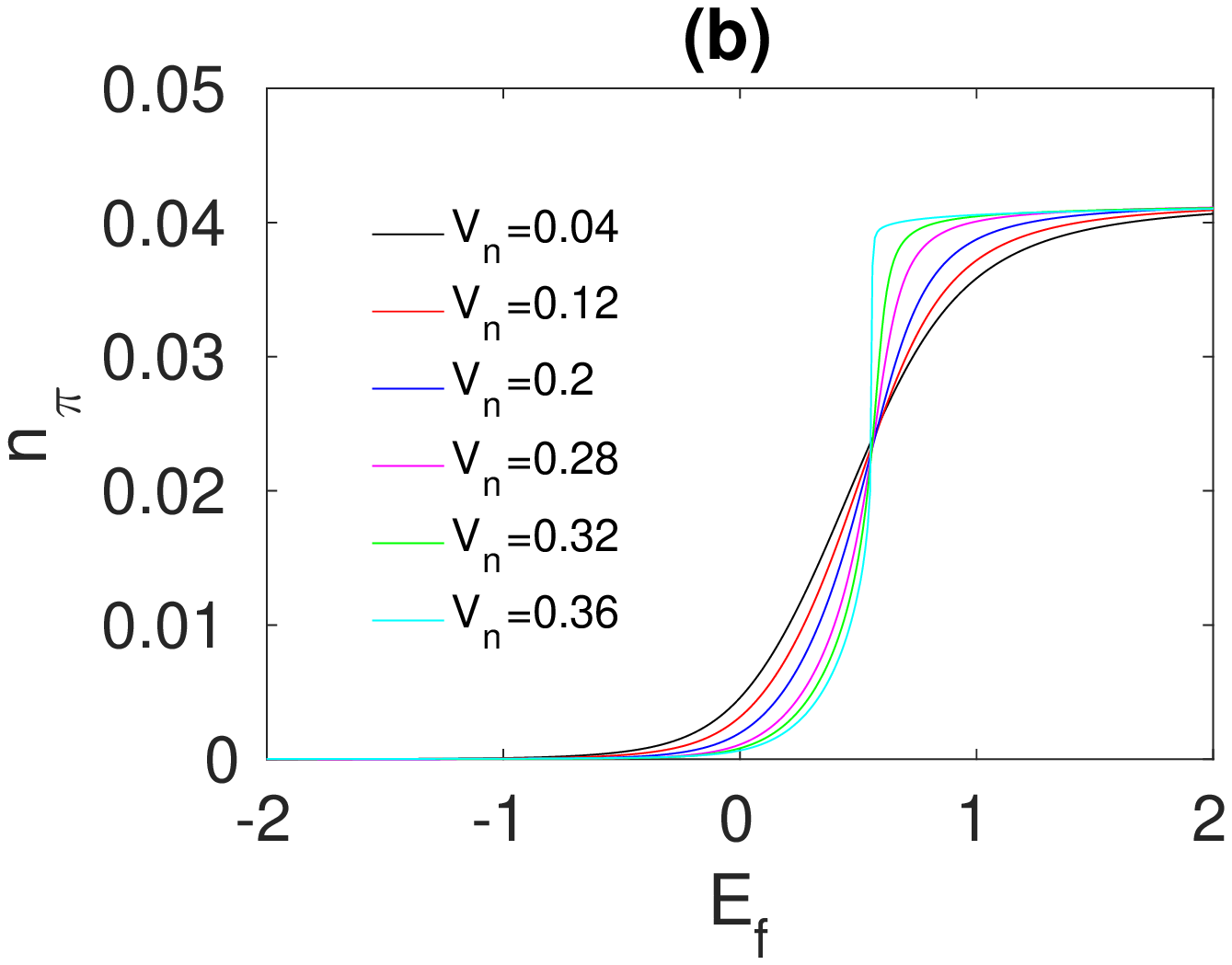}
\includegraphics[width=7.5cm]{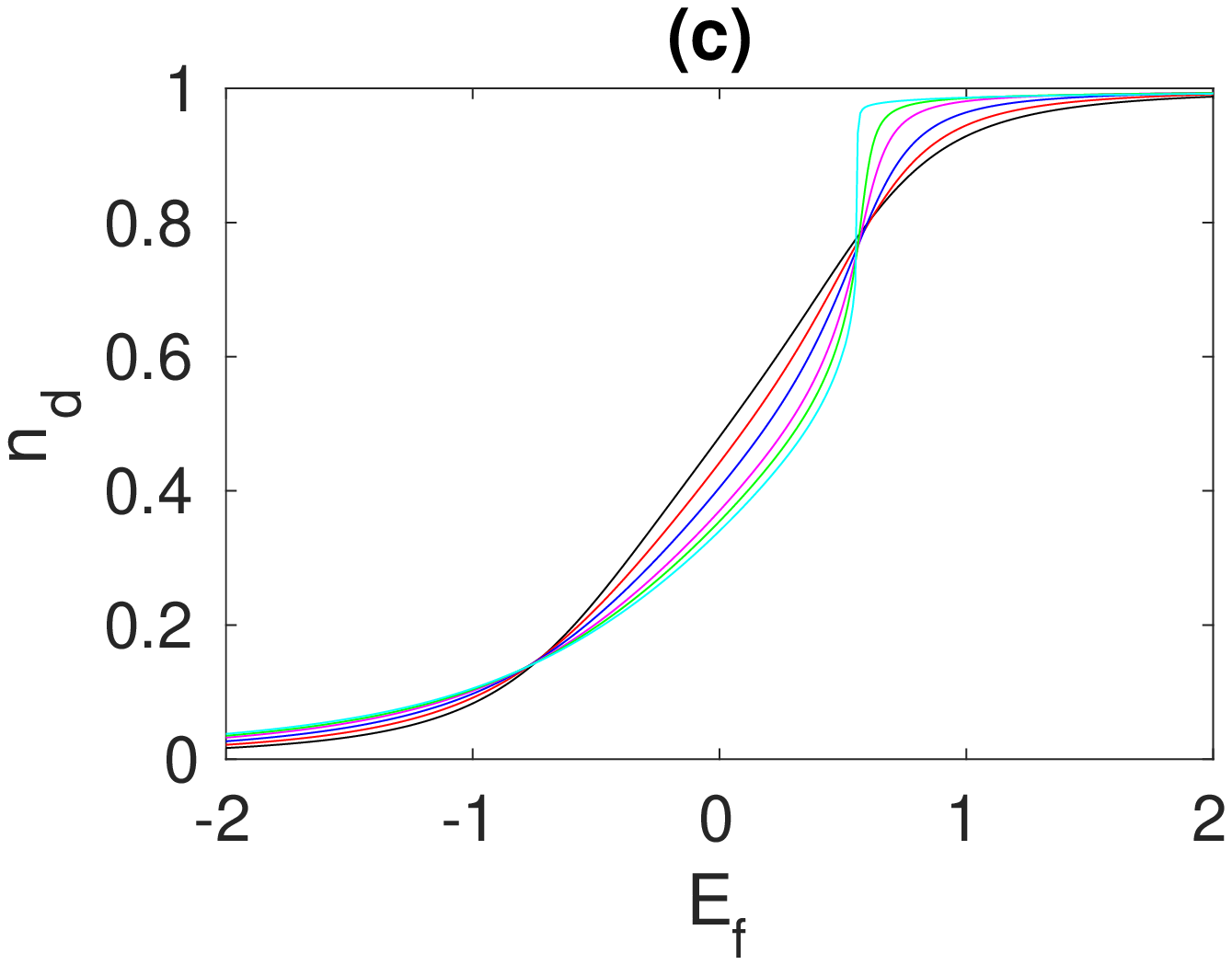}
\includegraphics[width=7.5cm]{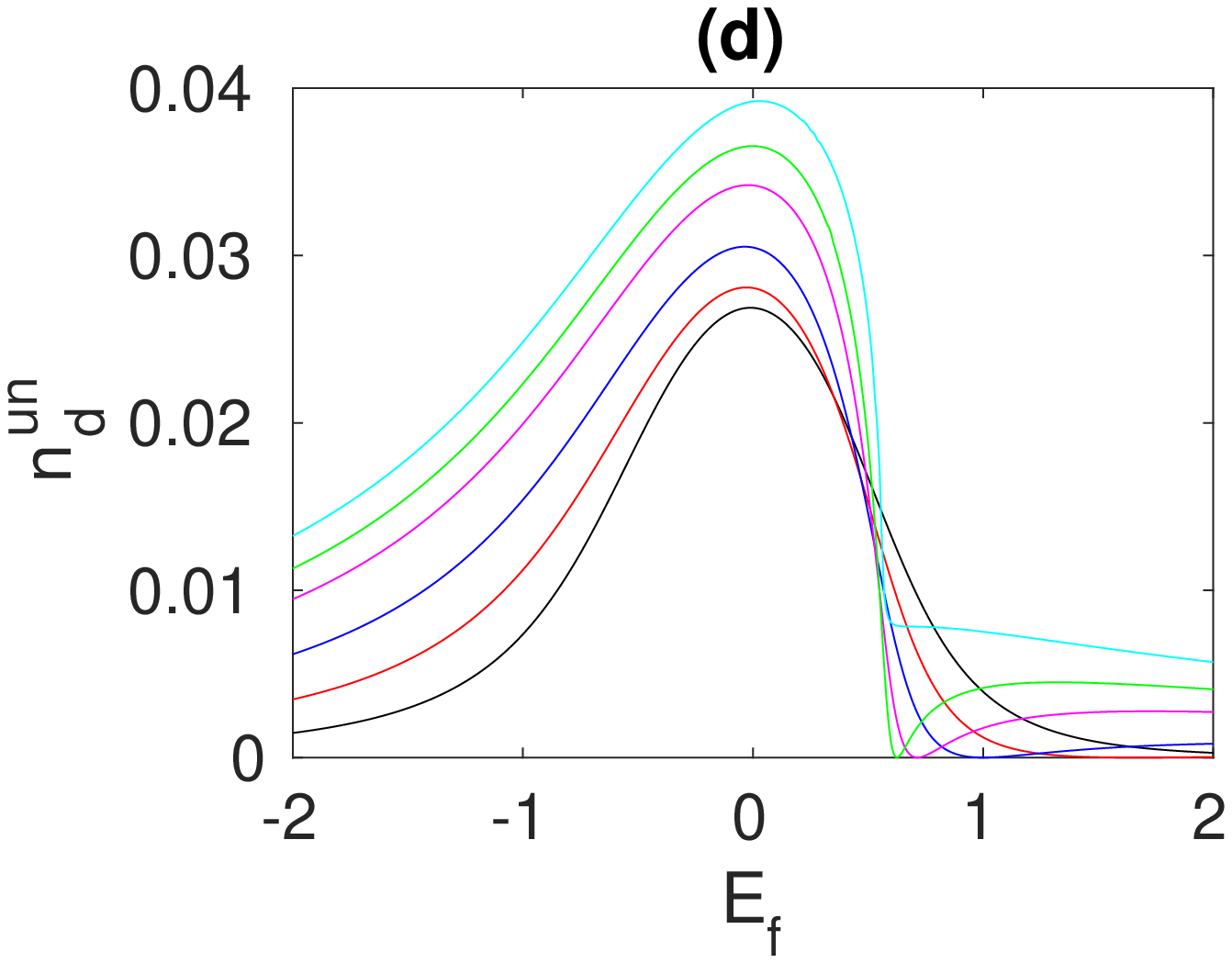}
\end{center}
\vspace*{-0.8cm}
\caption{\small The density of zero-momentum excitons $n_0$ (a),
the density of $\pi$-momentum excitons $n_{\pi}$ (b),
the total $d$-electron density $n_d$ (c), and the total density of unbound 
$d$ electrons $n^{un}_d=n_d-n_T$ (d) as functions of $E_f$ 
calculated for $U=4, V=0.2, L=60$ and six different values of $V_n$.}
\label{figure2}
\end{figure}
In all examined cases, the density of zero-momentum excitons is 
the most significantly enhanced for $d$-electron densities near 
the half-filled band case $E_f=0$ and $n_d = 1/2$. The changes of 
$n_0$ are gradual for $E_f<0$ and very steep, but still continuous,
for $E_f >0$. The fully different behaviour 
exhibits the density of $\pi$-momentum excitons $n_{\pi}$. Its
enhancement with increasing $E_f$ is practically negligible 
for $E_f < 0$, but from this value $n_{\pi}$ starts to increase 
sharply and tends to its saturation value corresponding to the
fully occupied $d$ band $n_d \sim 1$. The density of unbound $d$ 
electrons $n^{un}_d$  exhibits very
simple  behaviour for $E_f <0$. In this limit $n^{un}_d$ gradually increases 
with increasing $E_f$ for all examined values of nonlocal hybridization 
$V_n$. However, in the opposite case ($E_f > 0$) the density of unbound
$d$ electrons $n^{un}_d$  behaves fully differently for $V_n < V^c_n$
and $V_n > V^c_n$, where $V^c_n \sim 0.2$. For $V_n < V^c_n$ 
the density of unbound $d$ electrons $n^{un}_d$ gradually decreases 
with increasing $E_f$ and tends to zero when $E_f$ approaches the 
upper edge of the noninteracting band $E_f=2$,
but in the opposite limit the  density of unbound $d$ electrons 
$n^{un}_d$ decreases on the interval of $E_f$ values from $E_f=0$ to
$E^c_f(V_n)$, and $n^{un}_d$ starts to increases again for 
$E_f > E^c_f(V_n)$.
\begin{figure}[h!]
\begin{center}
\includegraphics[width=16cm]{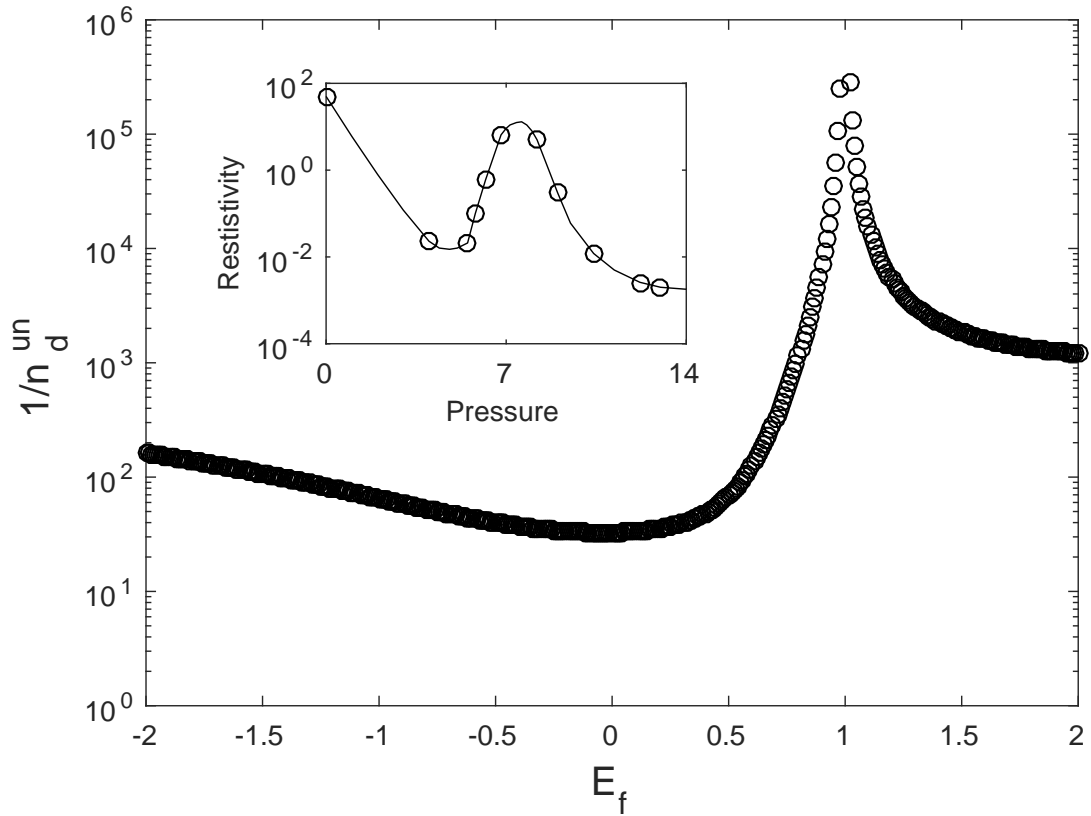}
\end{center}
\vspace*{-0.8cm}
\caption{\small The inverse value of the density of unbound $d$-electrons 
$n^{un}_d$ as a function of the $f$-level energy $E_f$ calculated for 
$U=4, V=0.2, V_n=0.2$ and $L=\infty$. The inset shows the resistivity 
as the function of pressure in $TmSe_{0.45}Te_{0.55}$ at
4.2~K~\cite{Wachter}.}
\label{figure2}
\end{figure}
Taking into account the above mentioned 
parametrization between $E_f$ and the external pressure $p$, as well as 
the fact that the electrical conductivity is proportional to the density 
of unbound electrons  $n^{un}_d$ (and the electrical resistivity to  $1/n^{un}_d$), 
the results discussed above could have very important physical 
consequences. Indeed, in Fig.~5 we have plotted the  quantity 
$1/n^{un}_d$ (in the logarithmic scale) as a function of $E_f$ and compare 
it with experimental measurements of the pressure dependence of the 
electrical resistivity in mixed valence compound $TmSe_{0.45}Te_{0.55}$
(see the inset in Fig.~5). One can see that there is a nice qualitative 
accordance between our theoretical predictions and experimental results 
of Wachter et al.~\cite{Wachter}. In spite of the fact that our model is in many
aspects very simplified, the physics that could lead to the unusual behaviour 
of the electrical resistivity in $TmSe_{0.45}Te_{0.55}$ under the external 
pressure seems to be clear. This is a result a  formation and condensation 
of excitonic bound states of conduction-band electrons and valence-band holes.

In summary, the combined effects of local and nonlocal hybridization on the 
formation and condensation of the excitonic bound states in the extended 
Falicov-Kimball model have been studied by DMRG method. 
The analysis of the  resultant behaviours of the  excitonic 
momentum distribution $N(q)$ showed that (i) the local
hybridization $V$  supports the formation of the ferroelectric 
$q=0$ momentum condensate and  the nonlocal hybridization $V_n$ supports 
the formation of the antiferroelectric $q=\pi$ momentum condensate, 
(ii) the combined effect of local and nonlocal hybridization further enhances 
the excitonic correlations in $q=0$ as well as $q=\pi$ state, especially 
for $V$ and $V_n$ values from the CDW region,  (ii) strong effects of local 
and nonlocal hybridization are observed also for other ground-state 
quantities of the model such as the  $f$-electron density, or the density 
of unbound $d$-electrons, which are generally enhanced with increasing 
$V$ and $V_n$ ($E_f=0$), (iv) the model can yield a reasonable explanation 
for the pressure-induced resistivity anomaly observed experimentally 
in $TmSe_{0.45}Te_{0.55}$ compound.

\vspace{0.5cm}
This work was supported by Slovak Research and Development Agency (APVV)
under Grant APVV-0097-12 and ERDF EU Grants under the contract No.
ITMS 26220120005 and ITMS26210120002.

\end{document}